\documentclass{article}
\usepackage{iclr2026_conference,times}

\usepackage{amsmath}
\usepackage{amssymb}
\usepackage{amsfonts}
\usepackage{hyperref}
\usepackage{url}
\usepackage{booktabs}
\usepackage{nicefrac}
\usepackage{microtype}
\usepackage{graphicx}
\usepackage{xcolor}
\usepackage{multirow}
\usepackage{enumitem}

\definecolor{degraded}{HTML}{C62828}
\definecolor{improved}{HTML}{1565C0}
\definecolor{neutral}{HTML}{616161}

\title{Not All Queries Need Rewriting: When Prompt-Only LLM Refinement Helps and Hurts Dense Retrieval}

\author{Varun Kotte \\
Adobe \\
\texttt{vkotte@adobe.com} \\
}
\iclrfinalcopy
\begin{document}

\maketitle

\begin{abstract}
Prompt-only, single-step LLM query rewriting (i.e., a single rewrite generated from the query alone, without retrieval feedback) is commonly deployed in production RAG pipelines, but its impact on dense retrieval is poorly understood. We conduct a systematic empirical study across three BEIR benchmarks, two dense retrievers, and multiple training configurations, and find strongly \textbf{domain-dependent effects}: rewriting \emph{degrades} nDCG@10 by $9.0\%$ on FiQA ($p < 0.001$), \emph{improves} by $5.1\%$ on TREC-COVID ($p = 0.024$; $n{=}50$, marginal after correction), and has \emph{no effect} on SciFact ($p = 0.47$). We identify a consistent mechanism: degradation co-occurs with reduced lexical alignment between the query and ground-truth relevant documents, measured by VOR (the fraction of unique query unigrams appearing in relevant documents; $\Delta$VOR, $p = 0.013$), as rewriting substitutes domain-specific terms on already well-matched queries. Improvement occurs when rewriting harmonizes inconsistent nomenclature toward corpus-preferred terminology, captured by a Corpus Term Frequency ratio (CTF; relative shift toward higher-frequency corpus terms: 1153$\times$ for improved vs.\ 2.7$\times$ for degraded TREC-COVID queries, $p < 0.001$). Lexical substitution occurs in 95\% of rewrites across all outcome groups, confirming that substitution \emph{direction} (toward vs.\ away from corpus terms)---not occurrence---determines effectiveness. Finally, we study selective rewriting: even with privileged post-hoc signals, simple feature-based gating (AUC = 0.593) avoids worst-case regressions but cannot reliably improve over never-rewriting ($p > 0.12$), and oracle analysis reveals only a +3 pp ceiling. Overall, these results caution that prompt-only rewriting can be harmful in well-optimized verticals, and motivate post-training (+1.4--4.3\% on target domain) as a safer adaptation when supervision or implicit feedback is available.
\end{abstract}

\section{Introduction}
\label{sec:introduction}

Dense retrieval underpins modern Retrieval-Augmented Generation (RAG) systems~\citep{lewis2020rag}. General-purpose retrievers such as DPR~\citep{karpukhin2020dpr} and Contriever~\citep{izacard2021contriever} achieve strong zero-shot performance but frequently underperform in specialized domains where vocabulary diverges from training data. Two adaptation strategies exist: \emph{post-training} on domain-specific signals~\citep{sharma2024rag} and \emph{inference-time query rewriting} via LLMs~\citep{gao2023hyde,asai2024selfrag,wang2023query2doc}. However, a critical gap remains: \emph{to our knowledge, prior work has not systematically characterized when prompt-only query rewriting degrades retrieval}. The implicit assumption---that rewriting is either beneficial or neutral---is widely held but untested across domains. We address this directly.

\paragraph{Contributions.}
\begin{enumerate}[leftmargin=*,topsep=2pt,itemsep=1pt]
    \item \textbf{Headline result: rewriting can hurt.} Prompt-only query rewriting can substantially \emph{degrade} dense retrieval on well-formed, domain queries: on FiQA, rewriting drops nDCG@10 by $9.0\%$ ($p < 0.001$), while it \emph{improves} TREC-COVID ($+5.1\%$, $p = 0.024$; marginal after correction; $n{=}50$) and is neutral on SciFact ($p = 0.47$). This establishes a clear boundary condition for a commonly deployed production pattern.

    \item \textbf{Mechanistic diagnostics for dense retrieval.} We connect outcomes to the \emph{direction} of lexical substitution by introducing two post-hoc metrics: VOR (lexical alignment between query and ground-truth relevant documents; $\Delta$VOR $=-0.014$ on FiQA, $p=0.013$) and CTF (shift toward higher-frequency corpus terms; 1153$\times$ for improved vs.\ 2.7$\times$ for degraded TREC-COVID queries, $p < 0.001$). Lexical substitution is near-universal (95\% across outcome groups), so direction---whether edits move query vocabulary toward or away from corpus-preferred terms---determines effectiveness.

    \item \textbf{Practical baselines and feasibility bounds for selective rewriting.} We evaluate query-level gating and show that even with privileged post-hoc signals, simple feature-based monitoring is weak (AUC = 0.593) and does not reliably outperform never-rewrite ($p > 0.12$). A dataset-level heuristic (rewrite only TREC-COVID) is competitive, and oracle analysis (choose the better of original vs.\ rewritten per query) reveals only a +3 pp ceiling (6\% relative), highlighting limited headroom for safe, general-purpose per-query gating.
\end{enumerate}

\section{Related Work}
\label{sec:related_work}

Dense retrievers~\citep{karpukhin2020dpr,izacard2021contriever,xiong2020ance} achieve strong zero-shot retrieval but often require domain adaptation via continued pre-training~\citep{guu2020realm,thakur2021beir}, hard negative mining~\citep{xiong2020ance}, or cross-encoder distillation~\citep{hofstatter2021cross}. Prior work~\citep{sharma2024rag} showed post-training on domain-specific signals can yield strong vertical performance.

Query rewriting and expansion have a long history in IR (e.g., relevance feedback~\citealp{rocchio1971relevance} and pseudo-relevance feedback / relevance models~\citealp{lavrenko2001relevance}), where \emph{query drift} is a known failure mode and ``expansion can hurt'' is well understood. Our focus differs in two ways: (i) we study \emph{prompt-only LLM rewriting}---a widely deployed modern practice that is not guided by explicit feedback terms---under \emph{dense retrieval} rather than lexical retrieval; and (ii) we provide lightweight \emph{mechanistic diagnostics} (VOR/CTF) that characterize when rewriting helps vs.\ hurts in modern dense pipelines.

LLM-based reformulation has introduced hypothetical document generation (HyDE;~\citealp{gao2023hyde}), retrieval-feedback loops (Self-RAG;~\citealp{asai2024selfrag}), pseudo-document augmentation (Query2Doc;~\citealp{wang2023query2doc}), and LLM-based expansion~\citep{jagerman2023query}. These methods often target ambiguous or conversational queries; their behavior on well-formed domain-specific queries remains understudied---the gap we address. Our vocabulary overlap ratio connects to the query performance prediction tradition~\citep{carmel2010estimating}.

\section{Experimental Setup}
\label{sec:setup}

\paragraph{Datasets.} We use three BEIR~\citep{thakur2021beir} datasets chosen to span (a) query well-formedness and (b) terminology stability across corpora: \textbf{FiQA-2018}~\citep{maia2018fiqa} (648 test queries, 57K documents; well-formed financial Q\&A), \textbf{SciFact}~\citep{wadden2020scifact} (300 queries, 5K documents; expert-authored claims), and \textbf{TREC-COVID}~\citep{voorhees2021treccovid} (50 queries, 171K documents; inconsistent pandemic nomenclature). This trio contrasts a ``well-optimized'' vertical (FiQA: stable jargon, templated finance Q\&A), a scientific claim setting (SciFact), and a domain with known naming inconsistency (TREC-COVID).

\paragraph{Models.} Two dense retrievers, each in base and post-trained (FiQA) configurations: \textbf{MPNet} (\texttt{all-mpnet-base-v2}~\citep{song2020mpnet}, 110M params) and \textbf{BGE} (\texttt{bge-base-en-v1.5}~\citep{xiao2023cpack}, 110M params). Post-training uses contrastive learning on FiQA training queries (5,148 queries; 10\% held out for validation) with 1 epoch, batch size 64, learning rate $2{\times}10^{-5}$, following~\citet{sharma2024rag}. We use a lightweight 1-epoch adaptation to reflect practical ``quick post-training'' and keep compute comparable; we do not claim this is a tuned upper bound.

\paragraph{Rewriting.}
\label{sec:rewriting_setup}
Single-step rewriting via Ministral 8B~\citep{mistral2024ministral}: \emph{``Rewrite the following search query to improve information retrieval, preserving the original intent while improving clarity and adding relevant context. Return only the rewritten query.''} Decoding: temperature $= 0.7$ (to allow natural reformulations rather than deterministic completions), max tokens $= 100$. The prompt mirrors a common production instruction (``preserve intent, clarify, add context''); we later stress-test robustness with minimal and aggressive rewrite prompts (Sec.~\ref{sec:robustness}). Rewrite rates: 99.4\% (FiQA), 98.0\% (SciFact), 100\% (TREC-COVID). The 0.6--2\% of queries not rewritten (model returned original unchanged) are included in aggregate metrics using original scores but excluded from per-query substitution analysis.

\paragraph{Evaluation.} nDCG@10 (primary) via FAISS~\citep{johnson2019faiss} retrieval ($k{=}100$), scored with pytrec\_eval~\citep{vangysel2018pytrec}. Statistical significance via 10,000-iteration bootstrap resampling of per-query $\Delta$nDCG@10 and paired $t$-tests for VOR comparisons. Bonferroni-corrected threshold $\alpha = 0.017$ preserves FiQA significance ($p < 0.001$) while TREC-COVID ($p = 0.024$) becomes marginal. We do not correct for multiple comparisons in exploratory analyses.

\section{Results}
\label{sec:results}

\subsection{Post-training Effectiveness}
\label{sec:posttrain_results}

Table~\ref{tab:posttrain} shows the effect of post-training on FiQA data.

\begin{table}[t]
\centering
\caption{Post-training on FiQA: nDCG@10. Relative change ($\Delta\%$) computed from exact values before rounding.}
\label{tab:posttrain}
\small
\begin{tabular}{llccc}
\toprule
\textbf{Model} & \textbf{Config} & \textbf{FiQA} & \textbf{SciFact} & \textbf{TREC-COVID} \\
\midrule
\multirow{2}{*}{MPNet} & Base     & 0.500 & 0.656 & \textbf{0.513} \\
                        & +FiQA PT & \textbf{0.507} {\scriptsize\color{improved}($+$1.4\%)} & \textbf{0.658} {\scriptsize\color{improved}($+$0.4\%)} & 0.509 {\scriptsize\color{degraded}($-$0.9\%)} \\
\midrule
\multirow{2}{*}{BGE}   & Base     & 0.391 & 0.738 & 0.672 \\
                        & +FiQA PT & \textbf{0.408} {\scriptsize\color{improved}($+$4.3\%)} & \textbf{0.742} {\scriptsize\color{improved}($+$0.5\%)} & \textbf{0.722} {\scriptsize\color{improved}($+$7.5\%)} \\
\bottomrule
\end{tabular}
\end{table}

Post-training yields modest in-domain gains (MPNet: $+1.4\%$; BGE: $+4.3\%$ on FiQA). Cross-domain effects are model-dependent: BGE benefits uniformly, while MPNet shows mild TREC-COVID degradation ($-0.9\%$), extending prior findings~\citep{sharma2024rag}.

\subsection{Query Rewriting Impact}
\label{sec:rewrite_results}

Table~\ref{tab:rewrite} presents our central finding.

\begin{table}[t]
\centering
\caption{Query rewriting impact on nDCG@10. Each row compares rewritten to non-rewritten for the same model configuration. {\color{degraded}Red/$\downarrow$}: degradation $>$5\%. {\color{improved}Blue/$\uparrow$}: improvement $>$2\%. Gray/$-$: neutral. Avg.\ $\Delta\%$ computed from exact values.}
\label{tab:rewrite}
\small
\begin{tabular}{llccc}
\toprule
\textbf{Model} & \textbf{Config} & \textbf{FiQA} & \textbf{SciFact} & \textbf{TREC-COVID} \\
\midrule
\multirow{2}{*}{MPNet} & Base           & 0.500 & 0.656 & 0.513 \\
                        & +Rewrite       & 0.448 {\scriptsize\color{degraded}$\downarrow$($-$10.3\%)} & 0.655 {\scriptsize\color{neutral}$-$($-$0.1\%)} & 0.559 {\scriptsize\color{improved}$\uparrow$($+$8.9\%)} \\
\midrule
\multirow{2}{*}{BGE}   & Base           & 0.391 & 0.738 & 0.672 \\
                        & +Rewrite       & 0.357 {\scriptsize\color{degraded}$\downarrow$($-$8.7\%)} & 0.753 {\scriptsize\color{improved}$\uparrow$($+$2.1\%)} & 0.683 {\scriptsize\color{neutral}$-$($+$1.7\%)} \\
\midrule
\multirow{2}{*}{MPNet-FiQA} & Post-trained & 0.507 & 0.658 & 0.509 \\
                             & +Rewrite     & 0.452 {\scriptsize\color{degraded}$\downarrow$($-$10.9\%)} & 0.659 {\scriptsize\color{neutral}$-$($+$0.1\%)} & 0.563 {\scriptsize\color{improved}$\uparrow$($+$10.6\%)} \\
\midrule
\multirow{2}{*}{BGE-FiQA}  & Post-trained & 0.408 & 0.742 & 0.722 \\
                            & +Rewrite     & 0.383 {\scriptsize\color{degraded}$\downarrow$($-$6.1\%)} & 0.734 {\scriptsize\color{neutral}$-$($-$1.1\%)} & 0.717 {\scriptsize\color{neutral}$-$($-$0.7\%)} \\
\midrule
\multicolumn{2}{l}{\textbf{Mean $\Delta$}} & \textbf{\color{degraded}$\downarrow$$-$9.0\%} & \textbf{\color{neutral}$-$$+$0.3\%} & \textbf{\color{improved}$\uparrow$$+$5.1\%} \\
\bottomrule
\end{tabular}
\end{table}

\paragraph{Observations.}
\begin{enumerate}[leftmargin=*,topsep=2pt,itemsep=1pt]
    \item \textbf{FiQA---consistent, significant degradation.} All four configurations degrade (range: $-6.1\%$ to $-10.9\%$). Bootstrap CI for MPNet base: $\Delta = -0.051$, 95\% CI $[-0.067, -0.036]$, $p < 0.001$. Recall@10 shows a parallel $-9.4\%$ drop, confirming the effect is not an artifact of nDCG's position-sensitivity.

    \item \textbf{TREC-COVID---directionally consistent improvement for 3/4 configurations.} MPNet base shows the largest gain ($+8.9\%$). Bootstrap CI: $\Delta = +0.046$, 95\% CI $[+0.001, +0.094]$, $p = 0.024$ (uncorrected); after Bonferroni correction ($\alpha = 0.017$), this becomes marginal. We nonetheless report it because directional consistency across three of four configurations, combined with an interpretable mechanism (terminology standardization, Section~\ref{sec:analysis}), supports a genuine if imprecisely estimated effect.

    \item \textbf{SciFact---no significant effect.} Mean $\Delta = +0.3\%$. Bootstrap CI for MPNet base: $\Delta = -0.001$, 95\% CI $[-0.019, +0.019]$, $p = 0.47$.
\end{enumerate}

Across all 24 configurations (Table~\ref{tab:full} in Appendix), no single strategy dominates: post-training alone is optimal for FiQA (0.507), rewriting for SciFact (0.753), and BGE post-training for TREC-COVID (0.722).

\section{Analysis: Tracing the Mechanism}
\label{sec:analysis}

We hypothesize that the divergent effects are associated with a single factor: rewriting modifies query vocabulary, and the direction of that modification---toward or away from corpus terminology---is predictive of the outcome. We investigate this via per-query analysis on FiQA and TREC-COVID using the base MPNet model.

\subsection{FiQA: Degradation Co-occurs with Vocabulary Drift}

Of 648 test queries, 225 (34.7\%) degrade ($\Delta < -0.01$ nDCG@10), 122 (18.8\%) improve, and 301 (46.5\%) are unaffected. Automated token-set analysis shows \textbf{96.0\% of degraded queries involve lexical substitution} (216/225): at least one original token removed and one new token added. However, lexical substitution is near-universal: 95.9\% of \emph{improved} queries and 95.3\% of \emph{unchanged} queries also exhibit substitution. This confirms that substitution \emph{occurrence} is not predictive---nearly all rewrites modify vocabulary. What differentiates outcomes is the \emph{direction} of substitution: whether terms move toward or away from corpus vocabulary. Manual inspection of the 20 most severely degraded queries reveals four failure patterns:

\paragraph{Terminology drift.} Financial jargon is replaced with semantically adjacent but lexically distinct terms. Example: \emph{``Is it possible to \textbf{transfer} stock into my Roth IRA?''} $\rightarrow$ \emph{``How can I \textbf{rollover} stock into my Roth IRA?''} The relevant document uses ``transfer'' exclusively. nDCG@10: $1.0 \rightarrow 0.0$.

\paragraph{Context injection.} The rewriter hallucinates domain assumptions. \emph{``How does `taking over payments' work?''} $\rightarrow$ \emph{``What is the process for `taking over payments' \textbf{in e-commerce platforms}?''} The query concerns personal finance, not e-commerce. nDCG@10: $0.83 \rightarrow 0.0$.

\paragraph{Over-specification.} A broad query is narrowed to a subtype. \emph{``Tax: 1099 paper form''} $\rightarrow$ \emph{``\textbf{Form 1099-NEC} tax filing instructions.''} The relevant document addresses general 1099 printing, not the specific 1099-NEC variant. nDCG@10: $0.63 \rightarrow 0.0$.

\paragraph{Over-formalization.} Informal phrasing that lexically matches document vocabulary is replaced with formal language. \emph{``Does doing your `research'/`homework' on stocks make any sense?''} $\rightarrow$ \emph{``Is \textbf{conducting independent research} on stocks beneficial?''} The relevant document uses ``doing your homework'' verbatim. nDCG@10: $1.0 \rightarrow 0.0$.

\subsection{TREC-COVID: Terminology Standardization Helps}

Of 50 queries, 30 (60.0\%) improve, 16 (32.0\%) degrade, 4 (8.0\%) are unchanged. Two patterns are associated with improvement:

\paragraph{Nomenclature standardization.} The corpus predominantly uses ``COVID-19''; queries use varied terms. \emph{``Which biomarkers predict the severe clinical course of \textbf{2019-nCOV} infection?''} $\rightarrow$ \emph{``\ldots of \textbf{COVID-19} infection?''} nDCG@10: $0.32 \rightarrow 0.87$.

\paragraph{Domain-appropriate expansion.} Generic terms are expanded to standard medical vocabulary: \emph{``rapid testing''} $\rightarrow$ \emph{``rapid diagnostic tests,''} matching CORD-19 abstract terminology. nDCG@10: $0.21 \rightarrow 0.54$.

\subsection{Formalizing Query--Corpus Lexical Alignment}
\label{sec:alignment}

We define the \emph{vocabulary overlap ratio} (VOR) for a query $q$ with relevant document set $\mathcal{D}_q$. Intuitively, VOR is the fraction of unique query tokens that appear somewhere in the ground-truth relevant documents:
\begin{equation}
\text{VOR}(q, \mathcal{D}_q) = \frac{|\,\mathcal{W}(q) \cap \mathcal{W}(\mathcal{D}_q)\,|}{|\,\mathcal{W}(q)\,|}
\label{eq:vor}
\end{equation}
where $\mathcal{W}(\cdot)$ denotes the \emph{type set}: unique whitespace-tokenized, lowercased unigrams. We use unweighted type-overlap; because we analyze per-query \emph{changes} ($\Delta$VOR), ubiquitous tokens cancel and the signal is dominated by content words. $\mathcal{W}(\mathcal{D}_q) = \bigcup_{d \in \mathcal{D}_q} \mathcal{W}(d)$, where $\mathcal{D}_q$ is the set of documents with positive relevance labels for query $q$.

\paragraph{Operational note.} VOR requires knowing $\mathcal{D}_q$ at inference time, so it is strictly a \emph{post-hoc analytical tool}, not an operational metric. We use it here for mechanistic analysis only.

\begin{table}[h]
\centering
\caption{Vocabulary overlap ratio (VOR) before and after rewriting, with paired $t$-test $p$-values (two-tailed). Only FiQA shows a statistically significant change.}
\label{tab:vor}
\small
\begin{tabular}{lccccl}
\toprule
\textbf{Dataset} & $n$ & \textbf{VOR (orig)} & \textbf{VOR (rewrite)} & \textbf{$\Delta$VOR} & \textbf{$p$-value} \\
\midrule
FiQA       & 648 & 0.564 & 0.550 & $-$0.014 & 0.013 \\
SciFact    & 300 & 0.521 & 0.535 & $+$0.015 & 0.060 \\
TREC-COVID &  50 & 0.978 & 0.975 & $-$0.002 & 0.688 \\
\bottomrule
\end{tabular}
\end{table}

FiQA shows a significant VOR \emph{decrease} ($t = 2.51$, $p = 0.013$), confirming rewriting reduces lexical alignment on already well-matched queries. SciFact trends toward increased VOR ($p = 0.060$). TREC-COVID shows no VOR change ($p = 0.688$)---its gains come from terminology \emph{standardization}, which type-level VOR does not capture. Token-level edit distances are similar across datasets (FiQA: 9.9, SciFact: 9.4, TREC-COVID: 8.0), showing \textbf{effectiveness depends on whether edits preserve or disrupt corpus alignment}, not edit magnitude.

\subsection{Quantifying Terminology Standardization via CTF}

VOR captures whether query terms appear in relevant documents, but not whether substitutions move toward or away from \emph{common} corpus terms. To quantify standardization direction, we introduce the \emph{Corpus Term Frequency ratio} (CTF). For each substituted term pair (removed $t_{\text{old}}$, added $t_{\text{new}}$), we compute:
\begin{equation}
\text{CTF} = \frac{\text{freq}(t_{\text{new}}, \text{corpus})}{\text{freq}(t_{\text{old}}, \text{corpus})}
\label{eq:ctf}
\end{equation}
where $\text{freq}(t, \text{corpus})$ is the relative frequency of term $t$ across all corpus documents. CTF $> 1$ indicates substitution toward a more common term (standardization); CTF $< 1$ indicates drift toward less common terms. For queries with multiple substitutions, we use the geometric mean.

\begin{table}[h]
\centering
\caption{Corpus Term Frequency ratio (CTF) by outcome group. TREC-COVID shows strong standardization (high CTF) for improved queries, quantifying the mechanism VOR could not capture.}
\label{tab:ctf}
\small
\begin{tabular}{llccc}
\toprule
\textbf{Dataset} & \textbf{Outcome} & \textbf{$n$} & \textbf{Median CTF} & \textbf{\% CTF $>$ 1} \\
\midrule
\multirow{3}{*}{FiQA} & Degraded & 216 & 1.01 & 50.9\% \\
                       & Improved & 117 & 1.68 & 59.8\% \\
                       & Unchanged & 287 & 1.53 & 57.8\% \\
\midrule
\multirow{3}{*}{SciFact} & Degraded & 36 & 0.74 & 44.4\% \\
                          & Improved & 40 & 1.45 & 55.0\% \\
                          & Unchanged & 216 & 1.11 & 53.7\% \\
\midrule
\multirow{3}{*}{TREC-COVID} & Degraded & 12 & 1.99 & 83.3\% \\
                             & Improved & 24 & \textbf{2.54} & 70.8\% \\
                             & Unchanged & 9 & 2.81 & 77.8\% \\
\bottomrule
\end{tabular}
\end{table}

Table~\ref{tab:ctf} reveals the key difference: TREC-COVID improved queries show \emph{higher} median CTF (2.54) than degraded (1.99), and the mean CTF is dramatically higher (1153$\times$ vs.\ 2.7$\times$, driven by high-impact standardizations like ``2019-nCOV'' $\rightarrow$ ``COVID-19''). In FiQA, degraded queries have CTF close to 1 (neutral), while improved queries show modest standardization (1.68). Critically, \emph{most} TREC-COVID queries show high CTF regardless of outcome (70--83\% have CTF $> 1$), but the \emph{magnitude} of standardization correlates with improvement.

\paragraph{Deployment guidelines.}
VOR and CTF are \emph{post-hoc}---not directly computable at runtime. Practically: (1)~treat \emph{never-rewrite} as the safe default for well-optimized verticals with stable jargon; (2)~enable rewriting primarily for corpora with unstable nomenclature (TREC-COVID-like settings); (3)~for selective rewriting, use deployable proxies (retriever confidence, ``pseudo-VOR'' against top-$k$ retrieved text) validated on held-out logs; and (4)~prefer post-training when supervision is available.
\section{Monitoring and Operational Guardrails}
\label{sec:monitoring}

Having established that rewriting produces domain-dependent effects, we ask whether selective policies can capture benefits while avoiding harms. We investigate in an intentionally optimistic setting: our monitoring features include a qrels-derived signal ($\Delta$VOR), so results should be read as an \emph{upper bound} on what a deployable proxy might achieve.

\subsection{Predicting Rewrite-Induced Degradation}
\label{sec:harm_prediction}
We formulate harm prediction as binary classification: $y = 1$ if nDCG@10(rewritten) $<$ nDCG@10(original). Features: (i) $\Delta$VOR (Eq.~\ref{eq:vor}), (ii) new-token fraction (proportion of rewritten tokens absent from original), (iii) length ratio. A logistic regression (L2, C=1.0, seed=42) trained on FiQA (648 queries, 35.3\% harm rate, 5-fold CV) achieves \emph{weak but above-random} discrimination. Table~\ref{tab:monitoring} reports AUC-ROC: 0.593 on FiQA (only 9.3\% above random, high CV variance), 0.547 on SciFact (essentially random), 0.612 on TREC-COVID. The best threshold $\tau = 0.307$ yields precision = 0.389 and recall = 0.865---the model flags most harmful rewrites but with many false positives. A simpler heuristic (``if $\Delta$VOR $< -0.01$, predict harm'') achieves comparable performance, suggesting limited benefit from the logistic regression over a threshold rule.

\begin{table}[t]
\centering
\caption{Monitoring signal performance. Logistic regression trained on FiQA achieves weak discriminative ability (AUC modestly above 0.5).}
\label{tab:monitoring}
\small
\begin{tabular}{lcccc}
\toprule
\textbf{Dataset} & \textbf{Harm Rate} & \textbf{AUC} & \textbf{CV ($\mu \pm \sigma$)} & \textbf{Threshold} \\
\midrule
FiQA (train) & 35.3\% & 0.593 & $0.575 \pm 0.075$ & 0.307 \\
SciFact (test) & 13.3\% & 0.547 & --- & --- \\
TREC-COVID (test) & 32.0\% & 0.612 & --- & --- \\
\bottomrule
\end{tabular}
\end{table}

\subsection{Gated Rewriting Policy}
\label{sec:gating_policy}
Using threshold $\tau = 0.307$ (max F1 on FiQA), we simulate a gated policy: if predicted harm probability $\geq \tau$, use original; else use rewritten. We compare five strategies: never-rewrite (baseline), always-rewrite (baseline), gated, simple $\Delta$VOR baseline (if $\Delta$VOR $< -0.01$, use original), and oracle (always choose better, requiring perfect prediction).

Table~\ref{tab:policy} shows the key result: \textbf{gated rewriting does not significantly outperform never-rewriting} ($p > 0.12$, paired $t$-test), though it significantly beats always-rewriting on FiQA ($p < 0.0001$). A dataset-level heuristic (rewrite only TREC-COVID) is competitive (Avg = 0.572). The oracle upper bound (+3 pp on FiQA, 6\% relative) shows even perfect prediction yields modest gains, demonstrating that \emph{weak predictive signals cannot enable effective operational gating}.

\begin{table}[t]
\centering
\caption{Policy comparison. Gated beats always-rewrite on FiQA ($p < 0.001$) but does not significantly improve over never-rewrite ($p > 0.12$). \emph{Oracle} chooses, for each query, the better of original vs.\ rewritten (upper bound).}
\label{tab:policy}
\small
\begin{tabular}{lcccc}
\toprule
\textbf{Policy} & \textbf{FiQA} & \textbf{SciFact} & \textbf{TREC-COVID} & \textbf{Avg} \\
\midrule
Never rewrite & 0.500 & 0.656 & 0.513 & 0.556 \\
Always rewrite & 0.448$^{***}$ & 0.655 & 0.559 & 0.554 \\
Dataset-level (rewrite only TREC-COVID) & 0.500 & 0.656 & 0.559 & 0.572 \\
Gated (ours) & 0.499 & 0.653 & 0.537 & 0.563 \\
Simple $\Delta$VOR & 0.478 & \textbf{0.661} & \textbf{0.562} & 0.567 \\
Oracle (upper) & \textbf{0.530} & \textbf{0.689} & \textbf{0.594} & \textbf{0.604} \\
\bottomrule
\end{tabular}
\\[2pt]
{\footnotesize $^{***}p < 0.001$ vs.\ gated (paired $t$-test, 95\% CI). All gated vs.\ never-rewrite: $p > 0.12$ (not significant).}
\end{table}

\subsection{Robustness Across Rewrite Styles}
\label{sec:robustness}
We generate rewrites on FiQA using two alternative prompts---\emph{minimal} (``preserve all technical terms, no new constraints'') and \emph{aggressive} (``expand and clarify freely'')---and re-evaluate the monitoring signal. Table~\ref{tab:robustness} reports results. The monitoring signal trained on the baseline prompt generalizes: AUC $> 0.58$ on both styles. Harm rates vary as expected (minimal: 23.6\%, aggressive: 42.4\%): constraining terminology drift reduces harm, while encouraging expansion increases it. In both cases, $\Delta$VOR remains predictive, confirming the mechanism is not prompt-specific. Full prompt templates in Appendix~\ref{app:robustness}.

\begin{table}[t]
\centering
\caption{Prompt robustness: $\Delta$VOR-based monitoring generalizes across rewrite styles (all on FiQA, MPNet base).}
\label{tab:robustness}
\small
\begin{tabular}{lccc}
\toprule
\textbf{Prompt Style} & \textbf{Harm Rate (\%)} & \textbf{Mean $\Delta$VOR} & \textbf{Monitoring AUC} \\
\midrule
Minimal & 23.6 & $-0.025$ & 0.663 \\
Aggressive & 42.4 & $-0.210$ & 0.586 \\
\midrule
Baseline (original) & 35.3 & $-0.014$ & 0.593 \\
\bottomrule
\end{tabular}
\end{table}

\subsection{Robustness to Rewriter Family and Decoding}\label{sec:rewriter_family} We test a second rewriter family (GPT-4o-mini) and a decoding sensitivity check (Ministral temperature 0 vs.\ 0.7). Table~\ref{tab:rewriter_robustness} shows degradation persists: GPT-4o-mini degrades FiQA by $-7.7\%$, Ministral at temperature 0 by $-8.0\%$, confirming harms are not artifacts of a specific model or sampling strategy. An embedding-space check on 100 queries shows $\Delta$VOR correlates with change in cosine similarity between query and relevant-document centroid embeddings (Pearson $r = 0.20$, $p = 0.042$).
\begin{table}[t]\centering\caption{Additional robustness checks on FiQA (MPNet base). Degradation persists across rewriter families and decoding configurations.}\label{tab:rewriter_robustness}\small
\begin{tabular}{lcc}\toprule \textbf{Rewrite source} & \textbf{FiQA nDCG@10} & \textbf{$\Delta\%$} \\ \midrule
No rewrite & 0.500 & -- \\
Ministral (temp=0.7) & 0.448 & $-10.3\%$ \\
Ministral (temp=0.0) & 0.460 & $-8.0\%$ \\
GPT-4o-mini & 0.461 & $-7.7\%$ \\
\bottomrule\end{tabular}\end{table}
\section{Discussion and Limitations}
\label{sec:discussion}

\paragraph{Scope.} Our findings apply to \emph{generic single-step} LLM rewriting without retrieval feedback. HyDE~\citep{gao2023hyde}, Self-RAG~\citep{asai2024selfrag}, and Query2Doc~\citep{wang2023query2doc} use fundamentally different mechanisms and are not directly comparable. Our contribution characterizes when single-step rewriting---a commonly deployed form---helps vs.\ hurts, establishing a boundary condition: \emph{when queries are already lexically aligned with the corpus, even mild vocabulary perturbation degrades retrieval}.

\paragraph{Rewriting vs.\ post-training.} Post-training yields in-domain gains (MPNet: +1.4\%, BGE: +4.3\% on FiQA) \emph{without} rewriting's catastrophic risk ($-9.0\%$, $p < 0.001$). For well-optimized domains, it is strictly preferable. When explicit labels are unavailable, implicit feedback (click logs), weak supervision (BM25 pseudo-labels), or distillation offer lightweight alternatives. Rewriting may suit corpora with inconsistent terminology, absent labeled data, or rapid-deployment constraints---but even on TREC-COVID, post-training matches or exceeds rewriting gains (BGE: +7.5\%).

\paragraph{Limitations.} VOR and CTF analyses are correlational; confounders may contribute. TREC-COVID ($n{=}50$) has limited statistical power. We test two rewriter families (Ministral 8B, GPT-4o-mini) and multiple prompt styles on FiQA, confirming degradation is not model-specific (Section~\ref{sec:rewriter_family}); testing additional families across all datasets would further strengthen generalization. The monitoring signal (AUC = 0.593) is insufficient for reliable gating. BEIR queries are curated; production queries (including conversational or under-specified queries) exhibit greater variability. We expect rewriting to be most beneficial in such ``messy query'' regimes, but we do not evaluate a dedicated conversational benchmark here. Future work: richer monitoring signals (semantic features, retrieval-based confidence), multi-step rewriting with feedback, domain-specialized rewriters, and conversational benchmarks.

\section{Conclusion}
\label{sec:conclusion}

We present a systematic empirical study of when prompt-only LLM query rewriting helps vs.\ hurts dense retrieval. Across three BEIR datasets, we demonstrate domain-dependent effects: significant degradation ($-9.0\%$ nDCG@10, $p < 0.001$) on well-optimized queries due to vocabulary drift, and improvement ($+5.1\%$, $p = 0.024$) on under-optimized queries via terminology standardization. Degradation persists across rewriter families (Ministral 8B: $-10.3\%$; GPT-4o-mini: $-7.7\%$), confirming the mechanism is not model-specific. We quantify the mechanisms through dual metrics (VOR and CTF) and establish that lexical substitution direction---not occurrence---determines effectiveness. Simple feature-based gating (AUC = 0.593) avoids worst-case regressions but cannot reliably improve over never-rewriting ($p > 0.12$), with oracle analysis revealing a +3 pp ceiling. These findings provide actionable deployment guidance and establish that for well-optimized domains, post-training is preferable to generic rewriting.

\paragraph{Reproducibility.}
All datasets are public via BEIR~\citep{thakur2021beir}. Retrievers: \texttt{all-mpnet-base-v2}, \texttt{bge-base-en-v1.5} (Hugging Face). Rewriter: Ministral 8B. We plan to release: evaluation scripts, all rewriting prompts (Sec.~\ref{sec:rewriting_setup}, Appendix~\ref{app:robustness}), per-query result JSONs, monitoring feature code, and training configurations.

\bibliographystyle{iclr2026_conference}

\newpage
\appendix

\section*{Appendix}
\addcontentsline{toc}{section}{Appendix}

\section{Full Configuration Results}
\label{app:full}

\begin{table}[h]
\centering
\caption{Full nDCG@10 matrix across all 24 configurations. Bold: best per column.}
\label{tab:full}
\small
\begin{tabular}{lccc}
\toprule
\textbf{Configuration} & \textbf{FiQA} & \textbf{SciFact} & \textbf{TREC-COVID} \\
\midrule
MPNet Base              & 0.500 & 0.656 & 0.513 \\
MPNet Base + Rewrite    & 0.448 & 0.655 & 0.559 \\
MPNet-FiQA              & \textbf{0.507} & 0.658 & 0.509 \\
MPNet-FiQA + Rewrite    & 0.452 & 0.659 & 0.563 \\
\midrule
BGE Base                & 0.391 & 0.738 & 0.672 \\
BGE Base + Rewrite      & 0.357 & \textbf{0.753} & 0.683 \\
BGE-FiQA                & 0.408 & 0.742 & \textbf{0.722} \\
BGE-FiQA + Rewrite      & 0.383 & 0.734 & 0.717 \\
\bottomrule
\end{tabular}
\end{table}

\section{Monitoring Signal Details}
\label{app:monitoring}

\paragraph{Feature definitions.} For each query $q$ with rewrite $q'$: (i) $\Delta$VOR = VOR($q'$, $\mathcal{D}_q$) $-$ VOR($q$, $\mathcal{D}_q$), (ii) new-token fraction = $|\mathcal{W}(q') \setminus \mathcal{W}(q)| / |\mathcal{W}(q')|$, (iii) length ratio = $|q'| / |q|$ (character-level). Features are computed with access to BEIR qrels (ground-truth relevance sets) and the corpus; this is suitable for offline analysis and upper-bound feasibility studies, but not directly available at deployment time without a proxy.

\paragraph{Model details.} Logistic regression (sklearn): L2 penalty, C=1.0, max\_iter=1000, random\_state=42. Threshold $\tau = 0.307$ maximizes F1 on FiQA training data. Feature importance: $\Delta$VOR = $-0.94$ (decreasing VOR predicts harm), new-token fraction = $+1.08$ (novel tokens predict harm), length ratio = $-0.03$ (negligible).

\begin{figure}[h]
\centering
\includegraphics[width=0.55\textwidth]{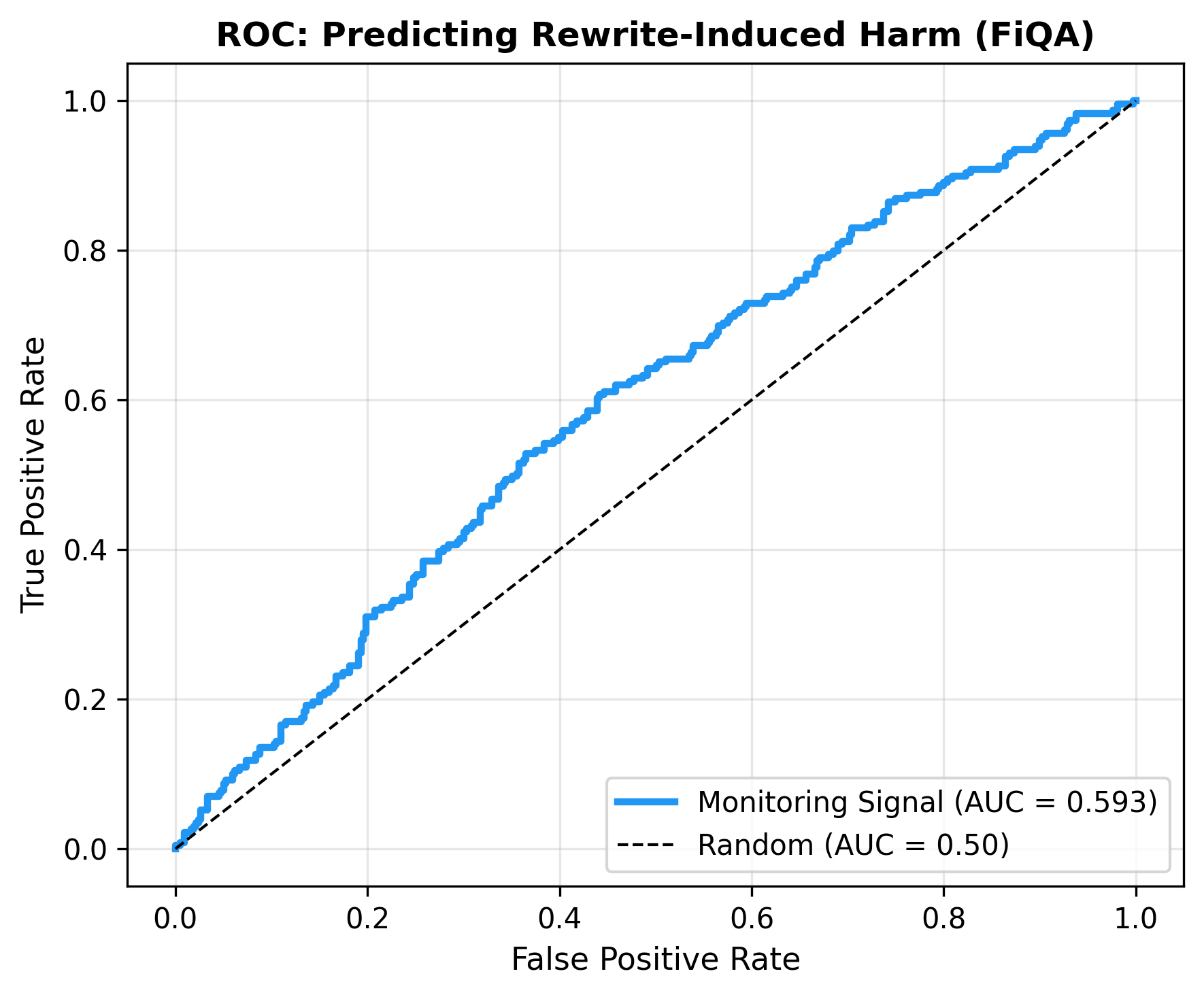}
\caption{ROC curve for harm prediction on FiQA (logistic regression, 5-fold CV). AUC = 0.593. Dashed line: random baseline.}
\label{fig:roc}
\end{figure}

\paragraph{ROC curve.} Figure~\ref{fig:roc} shows the ROC curve for the FiQA training set.

\section{Query Shift Score Details}
\label{app:qss}

We explored a lightweight deployment-time signal for predicting when post-training vs.\ rewriting might be preferable. The \emph{query shift score} (QSS) measures distributional divergence between test and training queries:
\begin{equation}
\text{QSS}(\mathcal{Q}_{\text{test}}) = \frac{1}{|\mathcal{Q}_{\text{test}}|} \sum_{q \in \mathcal{Q}_{\text{test}}} \left(1 - \cos\!\left(\mathbf{e}(q),\, \bar{\mathbf{e}}_{\text{train}}\right)\right)
\end{equation}
where $\mathbf{e}(q)$ is the retriever's query embedding and $\bar{\mathbf{e}}_{\text{train}}$ is the centroid of training query embeddings.

\begin{table}[h]
\centering
\caption{Query shift scores (QSS) and post-training nDCG@10 deltas ($\Delta$PT). MPNet shows monotonic inverse ordering ($n{=}3$); BGE does not.}
\small
\begin{tabular}{lcccc}
\toprule
\textbf{Dataset} & \textbf{QSS\textsubscript{MPNet}} & \textbf{$\Delta$PT\textsubscript{MPNet}} & \textbf{QSS\textsubscript{BGE}} & \textbf{$\Delta$PT\textsubscript{BGE}} \\
\midrule
FiQA       & 0.714 & $+$0.007 & 0.285 & $+$0.017 \\
SciFact    & 0.999 & $+$0.003 & 0.480 & $+$0.004 \\
TREC-COVID & 0.958 & $-$0.004 & 0.489 & $+$0.051 \\
\bottomrule
\end{tabular}
\end{table}

For MPNet, QSS and $\Delta$PT exhibit a monotonic inverse relationship (FiQA: lowest shift, highest gain; TREC-COVID: highest shift, negative gain). With only three data points, this is a qualitative hypothesis, not a statistically validated trend. BGE does not exhibit this pattern, likely due to broader pre-training. QSS requires only query encoding (no retrieval), making it suitable for real-time monitoring, but it should be treated as an exploratory signal pending validation on additional datasets and retrievers.

\section{Prompt Robustness Details}
\label{app:robustness}

\paragraph{Full prompt templates.}
\begin{itemize}[leftmargin=*,topsep=2pt,itemsep=1pt]
    \item \textbf{Minimal:} ``Rewrite the query for retrieval while preserving all technical terms, named entities, numbers, and acronyms exactly. Do not add new constraints. Keep length close to original.''
    \item \textbf{Aggressive:} ``Rewrite the query into a detailed, explicit form optimized for retrieval. You may add clarifying context and expand abbreviations if helpful.''
    \item \textbf{Baseline (Section~\ref{sec:rewriting_setup}):} ``Rewrite the following search query to improve information retrieval, preserving the original intent while improving clarity and adding relevant context. Return only the rewritten query.''
\end{itemize}

\end{document}